\documentclass[aps,pre,twocolumn,groupedaddress]{revtex4}

\usepackage{graphicx}
\usepackage[amssymb]{SIunits}             	
\usepackage[usenames,dvipsnames]{color}

\newcommand{\romannum}[1]{(\textit{\romannumeral #1})}

\begin{document}

\title{Interaction and coalescence of large bubbles rising in a thin gap}

\author{Sander G. Huisman$^{1,2}$, Patricia Ern$^{1,2}$ and V\'eronique Roig$^{1,2}$}
\email[]{roig@imft.fr}
\affiliation{$^1$ Universit\'e de Toulouse; INPT, UPS; IMFT (Institut de M\'ecanique des Fluides de Toulouse);  
All\'ee Camille Soula, F-31400 Toulouse, France}
\affiliation{$^2$ CNRS; IMFT; F-31400 Toulouse, France}

\date{15 February 2012}

\begin{abstract}
We present accurate measurements of the relative motion and deformation of two large bubbles released consecutively in a quiescent liquid confined in a thin-gap cell. Though the second bubble injected is smaller, we observed that in all cases it accelerates and catches up with the leading bubble. This acceleration is related to the wake of the leading bubble which also induces significant changes in the width and curvature of the trailing bubble. On the contrary, the velocity of the leading bubble is unaltered during the whole interaction and coalescence process. Shape adaptation of the two bubbles is observed just prior to coalescence. After pinch-off, the liquid film is drained at a constant velocity.

\end{abstract}

\maketitle

When two bubbles are rising in a quiescent fluid, the paths they follow is related to the perturbations they induce in the liquid and to their capacity to undergo deformation. In the tandem configuration (two bubbles rising along their line of centres), the wake of the leading bubble is a powerful mean to draw the bubbles together. This behavior is expected to promote coalescence, though escaping behavior may occur owing to the bubble's ability to deform or to lateral perturbations that may cause the bubbles to rotate and line up horizontally \cite{Hallez2011,Kok1993b}. Coalescence~of~a leading bubble with a trailing bubble is commonly observed in experimental high-Reynolds number bubbly flows \cite{Velez2011}, but is currently not predicted by theoretical and numerical works. Potential flow theory predicts that two bubbles rising in tandem will repel each other \cite{Harper1970,Kok1993a}, flow reversibility leading to the existence of a stagnation point between the two bubbles. Moreover, numerical simulations for non-deformable spherical bubbles at moderate-to-large Reynolds numbers predict that bubbles in line reach an equilibrium distance for which the repulsive potential effect balances the attractive viscous effect related to the vorticity produced at the bubble surface \cite{Yuan1994,Hallez2011}. 

There is presently little quantitative information on the bubble's kinematics during their interaction and on its relationship with the bubble's deformation. The aim of this paper is to obtain a simultaneous characterization of the kinematics and deformation of two bubbles rising in tandem, which is presently not available in the literature. For this purpose, we consider the interaction of two large bubbles rising freely in line in a vertical Hele-Shaw cell. The case of large bubbles is particularly interesting since their shape is preserved by coalescence. The use of a Hele-Shaw cell has the advantage to constrain the motion and the deformation of the bubbles to the plane of the cell. 

\begin{figure}[ht!]
\begin{center}
\includegraphics[width=0.42\textwidth]{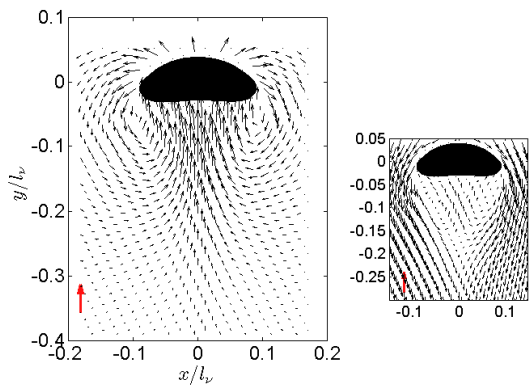}
\caption{(Color online) Liquid velocity around an isolated large bubble obtained by Particle Image Velocimetry (${\rm Ar}= 7000$, $V_{\infty}= 0.16 \, {\meter \per \second}$, $d=1.5 \, {\centi \meter}$). Left: in the laboratory frame; Right: focus on the recirculating wake in the bubble frame. The red thick arrow in both frames corresponds to $V_{\infty}$.}
\label{fig:PIV}
\end{center} 
\end{figure}

The motion of an isolated bubble rising in a quiescent liquid in a confined geometry has been studied in detail in Refs. \cite{Collins1965,Roig2011,Roudet2008,Tomiyama2006}. Given the fixed width $e$ of the cell, the equivalent diameter of the bubble is defined from the area $A$ of the bubble in the plane of the cell, $d=\sqrt{4A/\pi}$. Three non-dimensional parameters governing the single-bubble problem may then be introduced, the Archimedes number ${\rm Ar} = d\sqrt{gd}/\nu$, the Bond number ${\rm Bo}=\rho g d^2 / \sigma$ and the confinement ratio $e/d \ll 1$ ($g$ being the gravity acceleration, $\nu$ the liquid kinematic viscosity, $\rho$ the liquid density, and $\sigma$ the surface tension), the density and dynamic viscosity of the bubble being considered negligible relative to those of the liquid. For a given set of parameters, the resulting mean vertical velocity, $V_{\infty}$, of the bubble is used to define the Reynolds number ${\rm Re}=V_{\infty} \,d/\nu$. Furthermore, the dimensionless quantity $\tilde{{\rm Re}}={\rm Re}(e/d)^2$ compares the in-plane inertial timescale $V_{\infty}/d$ and the timescale for viscous diffusion in the gap $e^2/\nu$. The classical Hele-Shaw regime corresponds to $\tilde{{\rm Re}} \ll 1$ while inertial effects become predominant when $\tilde{{\rm Re}} \gg 1$ \cite{Bush1998}. In the present work, all the bubbles satisfy $6000<{\rm Ar}<18000$, ${\rm Re} \gg 1$ and $\tilde{{\rm Re}} \gg 1$ and have the shape of a circular segment. When these bubbles are rising alone, we measure that their terminal velocity is given by the relation ${\rm Re} = \sqrt{\frac{\pi}{2 C_d}} {\rm Ar} \simeq 0.42 {\rm Ar}$ or $V_{\infty} \simeq 0.42 \sqrt{g d}$, indicating that their drag coefficient is with a very good approximation independent of the bubble size ($C_d \simeq 3 \pi$), in agreement with \cite{Collins1965,Roig2011}. Upstream of the bubble, the flow can be conveniently predicted by potential flow theory whereas the production of vorticity at the bubble surface results in the existence of a stationary wake, as can be seen in fig. \ref{fig:PIV}. Moreover, the shear stresses at the walls impose a faster spatial decrease of the open wake than in an unbounded fluid, which is associated to the viscous lengthscale ${l}_{\nu} = V_{\infty} e^2/\nu$, namely $u / V_{\infty} \simeq \exp{(-10\,(y-y_R)/{l}_{\nu})}$,  where $y_R$ is the length of the recirculating wake measured from the bubble centre, $y_R \simeq 0.16 \, {l}_{\nu}$ \cite{Roig2011}. At a distance of about $0.4 \, {l}_{\nu}$ ($\simeq 4 \,d$ for ${\rm Ar} \simeq 7000$), the liquid velocity in the open wake is already $10\%$ of $V_{\infty}$. Note that ${l}_{\nu}/d \sim 1/\sqrt{d}$, so that as ${\rm Ar}$ increases the length of the recirculating wake decreases in terms of the bubble diameter ($ y_R \simeq 1.6 \, d$ for ${\rm Ar} \simeq 7000$ and $ \simeq 1.3 \, d$ for ${\rm Ar} \simeq 13500$).

The apparatus consists of two glass panes spaced \unit{1}{\milli \meter} apart and filled with water. The cell has a width of \unit{40}{\centi \meter} and a height of \unit{80}{\centi \meter}, allowing us to visualize the entire process of interaction and coalescence, while making sure the bubbles are not affected by end-effects. Gas injection is manually controlled by a combination of pressure reducing valves, stopcock valves, and dispensing needles. The bubbles are recorded using high-speed imaging, after which the background is subtracted from the frames. The resulting frames are binarized, revealing the contours of the bubbles. From the contours we obtain the kinematics and deformation characteristics of the bubbles. 

\begin{figure}[ht!]
\begin{center}
\includegraphics[width=0.35\textwidth]{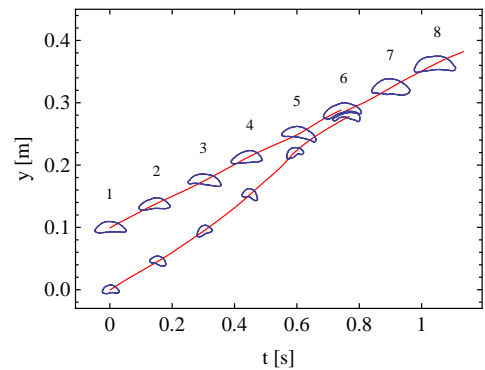}
\caption{(Color online) Overview of the trajectories of two coalescing bubbles (${\rm Ar_1} = 16700$, ${\rm Ar_2} = 8600$). The trailing bubble is entrained in the wake of the leading bubble (1-5), the bubbles adapt their shapes (6), and coalesce forming a single bubble (7, 8).}
\label{fig:Trajectory}
\end{center} 
\end{figure}

We vary the injected gas volume for both bubbles and release them quickly after one another from a single capillary. We investigated the case where the second bubble is smaller than the first one. If the delay between the bubbles injection is not too large, we observe that the trailing bubble catches up with the leading bubble, resulting in coalescence. Figure \ref{fig:Trajectory} provides an overview of this process. These experiments indicate that the suction effect provided by the wake of the leading bubble is sufficiently strong to accelerate the secondary bubble until it joins the first bubble. Moreover, this occurs even when the trailing bubble is smaller and has a smaller terminal velocity than the leading bubble, since the terminal velocity of an isolated bubble is proportional to $A^{1/4}$. Fig. \ref{fig:Trajectory} also reveals that a strong deformation of the secondary bubble occurs when it is sufficiently close to the leading bubble. We now provide a detailed characterization of the whole interaction and coalescence process, in particular the wake-induced relative motion of the two bubbles, their deformation when the bubbles are close enough and finally, the thin liquid film drainage occuring after pinch-off of the two bubbles interfaces. 

\begin{figure}[ht!]
 \begin{center}
 \includegraphics[width=0.4\textwidth]{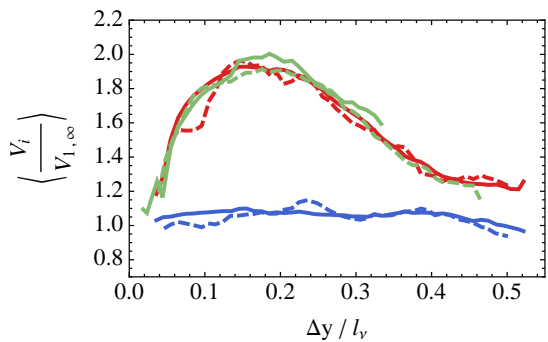}
  \caption{(Color online) Vertical velocity of the trailing bubble (top lines) and of the leading bubble (bottom lines), normalized by the terminal velocity of an isolated bubble of the same size as the leading bubble. $\langle \cdot \rangle$ denotes averaging over multiple experiments. Dark solid lines: results averaged over $40$ experiments, dark dashed lines: data from a single experiment (${\rm Ar_1} = 15447$, ${\rm Ar_2} = 8496$). The green (gray) curves compare two sets of experiments, both for $7500 < {\rm Ar_2} \leq 9000$ ($\langle {\rm Ar_2} \rangle = 8437$), and two different ranges of ${\rm Ar_1}$: dashed line for $11010 < {\rm Ar_1} < 12800$ ($\langle {\rm Ar_1} \rangle = 11806$) and solid line for $15950 < {\rm Ar_1} < 17520$ ($ \langle {\rm Ar_1} \rangle = 16685$).}
  \label{fig:Velocity}
 \end{center} 
\end{figure}

The data presented is the mean of 40 experiments, done over a range of ${\rm Ar}$ numbers ($10000<{\rm Ar_1}<17500$ and $6000<{\rm Ar_2}<10000$), where ${\rm Ar_2}/{\rm Ar_1}$ varies between $0.5$ and $0.8$ and where the indexes $1$ and $2$ correspond to the leading and trailing bubble, respectively. Fig. \ref{fig:Velocity} presents the instantaneous vertical velocities of the bubbles as a function of the distance between the centres of gravity of the bubbles. Both velocities are normalized with the terminal velocity of an isolated bubble of the same size as the leading bubble. On the abscissa, the distance between the centers of the bubbles is normalized using the viscous lengthscale $l_{\nu}$. Fig. \ref{fig:Velocity} shows that the speed of the leading bubble is unaltered by the presence of the second bubble. This has also been observed in Ref. \cite{Crabtree1971} for three-dimensional spherical caps, while for small spherical bubbles the velocity of both bubbles increases prior to coalescence \cite{Katz1996}. As regards the trailing bubble, fig. \ref{fig:Velocity} shows that it experiences a significant acceleration and then deceleration as it approaches the leading bubble. Thanks to the entrainment provided by the leading bubble's wake, the velocity of the trailing bubble is already $50\%$ higher than its isolated speed and $20\%$ higher than the speed of the leading bubble at $\Delta y \simeq 0.5 \, l_{\nu}$ (about four diameters away from the top bubble), and continues to increase while the trailing bubble is approaching the leading one. This acceleration phase was already measured in Ref. \cite{Crabtree1971} but they carried out no measurement beyond $\Delta y / d_1 \approx 2.5$. When the distance between the bubbles $\Delta y /  l_{\nu} \approx 0.2$ ($\Delta y / d_1 \approx 2$), the velocity of the trailing bubble reaches a maximum, about $2$ times faster than the speed of the leading bubble. This corresponds to frames 4 to 5 in fig. \ref{fig:Trajectory}. At that point, the trailing bubble is entering the attached recirculating wake of the leading bubble and it starts to experience large deformations, as will be described later. This velocity is kept until $\Delta y / l_{\nu} \approx 0.1$ ($\Delta y / d_1 \approx 1$). For lower separation distances, the liquid between the two bubbles is pushed out, the bubbles become wider, and the trailing bubble slows down and adapts its velocity to the speed of the leading bubble. Fig. \ref{fig:Velocity} also presents for trailing bubbles with ${\rm Ar_2} \approx 8437$ the evolution of their vertical velocity when the leading bubbles move either at ${\rm Ar_1} \approx 11806$ or at ${\rm Ar_1} \approx 16685$ (green curves). We observe that all the curves superpose satisfactorily. The same result is obtained when fixing ${\rm Ar_1}$ and considering two different sets of ${\rm Ar_2}$. This result indicates that, in the ranges of ${\rm Ar_1}$ and ${\rm Ar_2}$ investigated here, the evolution of the velocity of the trailing bubble with the separation distance follows a unique master curve, provided the velocity is normalized with the velocity of the leading bubble and the separation distance with the viscous lengthscale associated to the leading bubble. This result fully reveals the crucial role of the wake of the leading bubble in the interaction process. Note that, for comparison, data from a single experiment is also presented. This data is not as smooth as the averaged velocity, due to the shape oscillations of the body, but also in some cases due to the possible lack of alignment of the two bubbles and associated slight horizontal motion.

\begin{figure}[ht!]
 \begin{center}
 \includegraphics[width=0.33\textwidth]{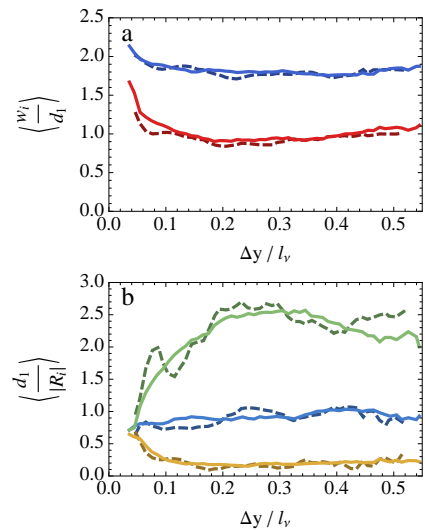}
 \caption{(Color online) (a): Evolution of the normalized width of the bubbles during approach. Top and bottom line correspond to the leading and trailing bubble, respectively. (b): Curvature of the bubbles as a function of separation distance. Top, middle, and bottom line correspond to the normalized curvature at the top of the trailing bubble, top of the leading bubble, and bottom of the leading bubble, respectively. Solid lines: results averaged over $40$ experiments, dashed lines: results from a single experiment (${\rm Ar_1} = 15447$, ${\rm Ar_2} = 8496$).}
  \label{fig:WidthCurvature}
 \end{center} 
\end{figure}

During the entire process the width and the curvature of the bubbles evolve. Fig. \ref{fig:WidthCurvature} shows the width $w$ and the curvature $1/R$ of the bubbles, normalized by the leading bubble diameter $d_1$, as a function of the separation distance. During the approach of the trailing bubble, from about 4 to 2 equivalent diameters apart (\textit{e.g.} frames 1-4 in fig. \ref{fig:Trajectory}), the top bubble does not change its shape significantly. The constant velocity of the first bubble is thus associated to a constant front curvature, as can be expected from potential flow theory \cite{Collins1965}. The trailing bubble, on the contrary, experiences a pronounced change in shape. It first becomes narrower, which causes the curvature to increase, as can be seen in fig. \ref{fig:WidthCurvature}. At $\Delta y / l_{\nu} \approx 0.2$ the width of the trailing bubble starts to increase, while its curvature and velocity start to decrease. As the bubble continues to approach the leading bubble, it becomes wider and slows down. In the final stage of approach the liquid between the two bubbles is pushed out, no immediate coalescence occurs and the leading bubble adjusts itself in order to accommodate for the trailing bubble by becoming slightly wider (fig. \ref{fig:WidthCurvature}). The bubbles adapt the shape of their interface and match each others curvature.

\begin{figure}[ht!]
 \begin{center}
  \includegraphics[width=0.4\textwidth]{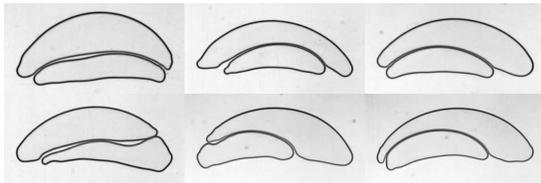}
  \caption{Overview of different pair of bubbles just prior to coalescence showing how the bubbles adapt their shape to each other. Note that each pair of bubbles is shaped like a single big bubble (${\rm Ar_1} \approx 11000$ and ${\rm Ar_2} \approx 7000$).}
  \label{fig:Precoalescence}
 \end{center} 
\end{figure}

Fig. \ref{fig:Precoalescence} shows an overview of the bubbles just before they coalesce. Most of the liquid between the bubbles has been drained, and the shape of the two bubbles together closely resembles a single large bubble. In this situation, the drag of the pair is very close to the leading bubble drag ($C_d \simeq 3 \pi$ and comparable velocities), while the drag of the pair is larger as long as the two bubbles have separated wakes. In addition, note that the bubbles are not always vertically aligned, despite being injected through the same nozzle, but in all cases the trailing buble is captured by the leading one.

\begin{figure}[ht!]
 \begin{center}
  \includegraphics[width=0.4\textwidth]{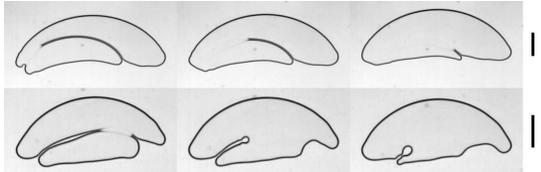}
  \caption{Two sequences of film drainage. Top three images: the liquid ligament is retracting at a constant velocity. Bottom images: the film changes geometry half-way and the fluid is collected in a rim. The scales on the right equal \unit{1}{\centi \meter}.}
  \label{fig:Drain}
 \end{center} 
\end{figure}

The final stage of coalescence is the merge of the interfaces; the fluid between the two bubbles has thinned enough so that non-hydrodynamical forces trigger the film rupture \cite{Probstein}. The drainage has been observed to be in several forms, depending on the geometry of the bubbles, as can be seen in fig. \ref{fig:Drain}. Film drainage can start at one end and then move across the entire bubble (see fig. \ref{fig:Drain}) or start in the middle upon which the fluid is evacuated in two directions. It is still an open question whether the fluid is partly drained in the films in between the bubble and the glass or in the sole plane of the cell. A thin liquid film exists between the moving bubbles and the walls. The thickness of this liquid film can be estimated by $h/e \approx Ca^{2/3} / (1 + Ca^{2/3})$ \cite{Roig2011,Aussillous2000}, where $Ca = \rho \nu V_{\infty}/\sigma$ is the capillary number. This gives, for the largest bubbles of our experiments, a film thickness of $h=\unit{25}{\micro \meter}$. In the experiments we used demineralised water, and the scaling law giving the velocity of the isolated bubble (${\rm Re}=0.42{\rm Ar}$) proves that the liquid speed in the films is negligible, indicating that a shear-free condition applies at the air-water interface of the thin films \cite{Collins1965,Roig2011}. When the film separating the two bubbles is uniform (absence of a rim) as in the upper sequence of fig. \ref{fig:Drain}, we found the speed of the moving interface in the frame of reference of the bubble and along the curved film to be constant and about $v_d \simeq \unit{1.65}{\meter \per \second}$. This measure was obtained with a time step of $\unit{0.5}{\milli \second}$ over a length of the retracting film of $\unit{3}{\centi \meter}$. Assuming that the drainage occurs in the plane of the cell and that the motion in the thin film is of Darcy type, {\it i.e.} that it results from a balance between pressure gradient and viscous friction at the walls, the mean drainage velocity in the gap (thickness $e$) is given by $v_d = \frac{e^2}{12 \rho \nu} \frac{\partial P}{\partial s}$, $s$ being the curvilinear abscissa along the film. The pressure gradient in the liquid film $\frac{\partial P}{\partial s}$ can be related to the difference in curvature along the 3D interface using Laplace's law. Considering that in the gap, the bubble curvature is fixed everywhere by the thickness $e$, the pressure gradient is at leading order driven by the change in the plane of the cell (over the lengthscale $d$) of the bubble curvature due to the local merging. An order of magnitude of $\frac{\partial P}{\partial s}$ is thus $\frac{1}{d} \frac{\sigma }{w}$, where the film thickness $w$ is smaller than the radius of curvature of the bubble at the film exit ($w \ll d$, see fig. \ref{fig:Drain}). This leads to $v_d \simeq \frac{\sigma}{12 \, \rho \nu} \frac{e^2}{w \, d}$, which is $O(1) \, {\meter \per \second}$ for $w \approx e$. A specific investigation should be carried out in the future to measure accurately the 3D geometry of the film and to verify that inertial effects can be neglected.  

Accurate simultaneous measurements of the kinematics and of the deformation of two large bubbles rising in line allowed us to identify and characterize the following different stages of interaction of the bubbles: \romannum{1} the trailing bubble is accelerated by the long-range wake of the primary bubble, \romannum{2} the second bubble enters the recirculating wake of the leading bubble, undergoes horizontal contraction and decelerates, \romannum{3} the bubbles adapt their shape to each other, in particular their curvatures are significantly modified, \romannum{4} the shape of both bubbles together resembles a single bubble and the liquid between the bubbles is squeezed out, and \romannum{5} the liquid film breaks and the bubbles merge. The quantitative results presented here for each stage may provide a valuable test for future theoretical and computational works on the road to predict bubbles attraction and coalescence.

We thank M. Roudet for performing the PIV measurement and S. Cazin for technical support concerning high-speed imaging. We also thank the E.U. Erasmus program, D. Lohse (University of Twente, The Netherlands) and D. Legendre (INPT, France) for the training grant allocated to S.G.H.

\end{document}